\documentclass{aa}           
\input{psfig.sty}

\begin{document}

\title{BeppoSAX observations of the three Gamma-ray pulsars PSR~B0656$+$14, 
PSR~B1055$-$52 and PSR~B1706$-$44}

\author{T. Mineo\inst{1}, E.Massaro\inst{2,3}, G.Cusumano\inst{1}, W.Becker\inst{4}}

\institute{Istituto di Astrofisica Spaziale e Fisica 
Cosmica, CNR, Sezione di Palermo, Via Ugo La Malfa 153, I-90146, Palermo,
Italy \and 
Dipartimento di Fisica, Universit\'a La Sapienza, Piazzale A. Moro 2,
I-00185, Roma, Italy \and
Istituto di Astrofisica Spaziale e Fisica Cosmica, CNR, 
Sezione di Roma, Via Fosso del Cavaliere,
I-00113, Roma, Italy \and
Max-Planck-Institut f\"ur Extraterrestrische Physik, D-85740, Garching-bei-M\"unchen,
Germany}

\offprints{T. Mineo: mineo@pa.iasf.cnr.it}

\date{Received:.; accepted:.}

\titlerunning{BeppoSAX observations of Gamma-ray pulsars}
\authorrunning{T. Mineo et al.}

\abstract{
We report the results of the observations of the three $\gamma$-ray pulsars 
PSR~B0656$+$14, PSR~B1055$-$52 and PSR~B1706$-$44 performed with BeppoSAX. 
We detected a pulsed emission only for PSR~B1055$-$52: in the range 0.1 
-- 6.5 keV the pulse profile is sinusoidal and the statistical significance 
is $4.5 \sigma$. The pulsed fraction was estimated 0.64$\pm$0.17.
This pulsation was detected also at energies greater than 2.5 keV suggesting 
either a non-thermal origin or a quite high temperature region on the
neutron star surface.  
Spectral analysis showed that only the X-ray spectrum of PSR~B1706$-$44 can be
fitted by a single power--law component, while that of PSR~B1055$-$52 requires
also a blackbody component ($kT = 0.075$ keV) and that of PSR~B0656$+$14 two
blackbody components  ($kT_1 = 0.059$, $kT_2 = 0.12$ keV).
}

\maketitle

\section{Introduction}

The study of the X-ray emission from $\gamma$-ray pulsars is important 
to gain more information about the acceleration site of particles and the 
high energy emission processes in the magnetosphere of these sources. 
Several models predict the production of a huge number of secondary 
electron-positron pairs generated by the interactions of primary high energy 
curvature photons with the intense magnetic field. These pairs are expected 
to radiate X rays via synchrotron and inverse Compton mechanisms.
This is not the only important emission processes of X-ray photons: thermal 
(blackbody) radiation from the polar caps, likely heated by the impinging of 
high energy particles, can also be quite efficient in this energy band.
Finally, in the case of young pulsars, X rays can be emitted in a compact 
synchrotron nebula surrounding the neutron star.
Several spectral components are then expected, likely characterized by different 
time modulation properties.

In this work we present the results of the spectral analysis in the energy
interval 0.1 to 10 keV of three pulsars PSR~B0656$+$14, PSR~B1055$-$52 and 
PSR~B1706$-$44, detected in the $\gamma$ rays by EGRET--CGRO, and observed by 
the Italian-Dutch satellite BeppoSAX. Only for one of them (PSR~B1055$-$52) we 
were able to detected pulsation, while for the other two a non modulated signal
was observed. 

PSR~B0656$+$14 ($P=0.384$ s, $\tau=P/2 \dot P=1.1 \times 10^5$ yr) was discovered
by Manchester et al. (1978) and its pulsed optical emission was detected by 
Shearer et al. (1997) and Pavlov, Welty \& Cordova (1997). 
In the X-ray band a pulsed signal at the radio period was observed by Cordova 
et al. (1989) and $\gamma$-ray detection in the EGRET band has been reported by 
Ramanamurthy et al. (1996).
ROSAT-PSPC observations (Finley et al. 1992) showed that 
the 0.1--2.4 keV spectral distribution can be well fitted  by two blackbodies 
or by a blackbody plus a power law. Greiveldinger et al. (1996), 
on the basis of ASCA and ROSAT data, proposed a three component model 
(two blackbodies plus a power law), while Wang et al. (1998) found that only 
two blackbodies without a power law are sufficient to have a satisfactory 
spectral fit.  

The middle-aged pulsar PSR~B1055$-$52 ($P=0.197$ s, $\tau=5.3\times10^5$ yr)
was discovered by Vaughan \& Large (1972). X-ray emission was first observed
by Chen \& Helfand (1983) with the {\it Einstein} Observatory and subsequently
by Brinkmann \& \"Ogelman (1987) with EXOSAT and by \"Ogelman \& Finley
(1993) with ROSAT who detected a nearly sinusoidal pulsation at energies 
up to 2.4 keV. 
The X-ray spectrum is complex and at least two components are necessary to
achieve an acceptable fit: one is a blackbody while the other is not safely
established and can be either a power law or another blackbody with  higher 
temperature.
Greiveldinger et al. (1996) prefer the two blackbody spectrum, with temperatures
of $7.9\times10^5$ K and $3.7\times10^6$ K; the power-law component would have
a photon index of 3--4 larger than the values measured in other pulsars. At variance,
Wang et al. (1998) give a flatter power-law fit with an index of 1.5$\pm$0.3.    
The pulsed emission above 100 MeV was first reported by Fierro et al. (1993) and
a comprehensive analysis of its high energy emission is given by Thompson et al. 
(1999). The pulse profile shows two peaks with  phase separation of 0.2 and
the spectrum is rather hard with a photon index of 1.58$\pm$0.15 below 1 GeV.
A study of the surrounding field and sources observed by BeppoSAX, 
ASCA and ROSAT has been recently presented by Becker et al. (1999) who concluded
against evidence for a wind--nebula around the pulsar.

PSR~B1706$-$44 is a young pulsar with a period of 0.102 s and a spin-down age of
$1.75\times10^4$ yr. It was discovered in the radio band by Johnston et al. (1992) 
and  unpulsed X-ray was first detected by Becker et al. (1995) with ROSAT-PSPC. 
In the more recent X-ray observations performed with 
ASCA (SIS+GIS) 
(Finley et al. 1998) and RossiXTE (Ray, Harding \& Strickman 1999) no pulsed 
emission was detected. SIS and GIS spectra in the (0.5--5) keV range were fitted 
by a power law with a photon index ranging from 1.6 to 1.9 and column densities 
of (1.3--2.2) $\times$ 10$^{21}$ cm$^{-2}$, but these values are poorly constrained because 
of their quite large (1 standard deviation) uncertainties of about 0.3 and 1.3 
$\times$ 10$^{21}$ (and even more), respectively. 
A spectral fit with the higher column density fixed at  5 $\times$ 10$^{21}$ 
cm$^{-2 
}$ gave the steeper photon index of 2.3 $\pm$ 0.3. 
A $\gamma$-ray source was detected by COS B (Swanenburg et al. 1981) 
and the pulsation at energies greater than  about 50 MeV was found by 
EGRET-CGRO (Thompson et al. 1992). An unpulsed 
source at TeV energies has been detected with the air Cherenkov telescope of 
the CANGAROO collaboration (Kifune et al., 1995).
VLA images by Frail et al. (1994) indicated that this 
pulsar may be located inside a plerionic nebula; new VLA images 
(Giacani et al. 2001) clearly shows that the pulsar is surrounded by a
syncrotron nebula about 3$''$.5 x 2$''$.5 in size.
Evidence for a X-ray compact nebula (with a radius of about 27") was also found 
by Finley et al. (1998) from the analysis of a ROSAT-HRI image. 
Very recently, using a Chandra-HRC-I observation Gotthelf, Halpern \& Dodson (2002) 
have discovered a pulsed X-ray emission from PSR~B1706$-$44 with a
nearly sinusoidal pulse profile and a pulsed fraction 23\%$\pm$6\%. 
Their spatial analysis provides a clear evidence for an extended
emission between 1$''$.5 and 20$''$ originated in a synchrotron nebula.

\medskip

\section{Observation and Data reduction}
The Italian-Dutch X-ray satellite BeppoSAX observed PSR~B0656$+$14 and 
PSR~B1706$-$44 in 1999 while the observation of PSR~B1055$-$52 was performed 
in December 1996, when all the three MECS units were working. 
Table 1 shows the observation log for the three pulsars together with the
LECS and MECS exposures.

\begin{table}
\label{tab1}
\caption{BeppoSAX observation log.}
\begin{tabular}{lccllccc}
\hline
\multicolumn{1}{c}{Source}  &  & \multicolumn{3}{c}{Date} &  & \multicolumn{2}{c}{Exposure (s)} \\
            &  &     &  &                &   & LECS    & MECS \\
PSR~B0656$+$14    &  & 9-11 & Mar & 1999    & & 34887.5 & 76002.9  \\
PSR~B1055$-$52     & & 28-29 & Dic & 1996   &  & 21818.5 & 55934.6\\
PSR~B1706$-$44     & & 29-31 & Mar & 1999    & & 36676.6 & 85554.5 \\
\hline
\end{tabular}
\end{table}

Standard procedures and selection criteria were applied to the data to avoid 
the South Atlantic Anomaly, solar, bright Earth and particle contamination 
using the SAXDAS v. 2.0.0 package. The images in the LECS and MECS 
show sources at positions fully compatible with the radio coordinates.

In the case of PSR~B0656$+$14  and PSR~B1055$-$52,  the events for the time and 
spectral analysis were selected within circular regions, centred at the radio 
position, with radii of  4' and 3' for the LECS and MECS, respectively.
These regions contain the 60\% of the 0.1--2.0 keV flux in LECS and the 82\% of 
the point source signal in the MECS; larger regions would include a higher 
background signal lowering the S/N ratio. 
The background was estimated from annular regions in the same fields and from 
a collection of bank field images.
As it will discussed later,  the spectrum of PSR~B0656$+$14 resulted quite 
complex and to obtain a more accurate estimate of the various components 
we need a significant increase of the statistics of the data. We considered, 
therefore, another observation of this pulsar, performed by ASCA on 1998 
October 11 and available from the archive, and joined it to our BeppoSAX 
data. We used only GIS data for which, after standard selections,
the observation time was 153 ks.   The spectrum was selected within a 
circular region of 3 arcmin and the background was accumulated in a 
source-free region of the field.

The event selection and the local background evaluation for PSR~B1706$-$44 
was not simple because of the presence of the near bright LMXB 
4U~1705$-$44, which lies close to the edge of the MECS, and just outside the LECS,
instrumental field of view. In order to get a reliable evaluation of the local 
background in the MECS image, we computed the count levels in a series of 
adjacent small circular regions with a radius of 3 arcmin   
located along the binary-to-pulsar direction (Fig. 1, 
upper panel).
 
\begin{figure}
\label{bkg}
\centerline{ 
\vbox{
\psfig{figure=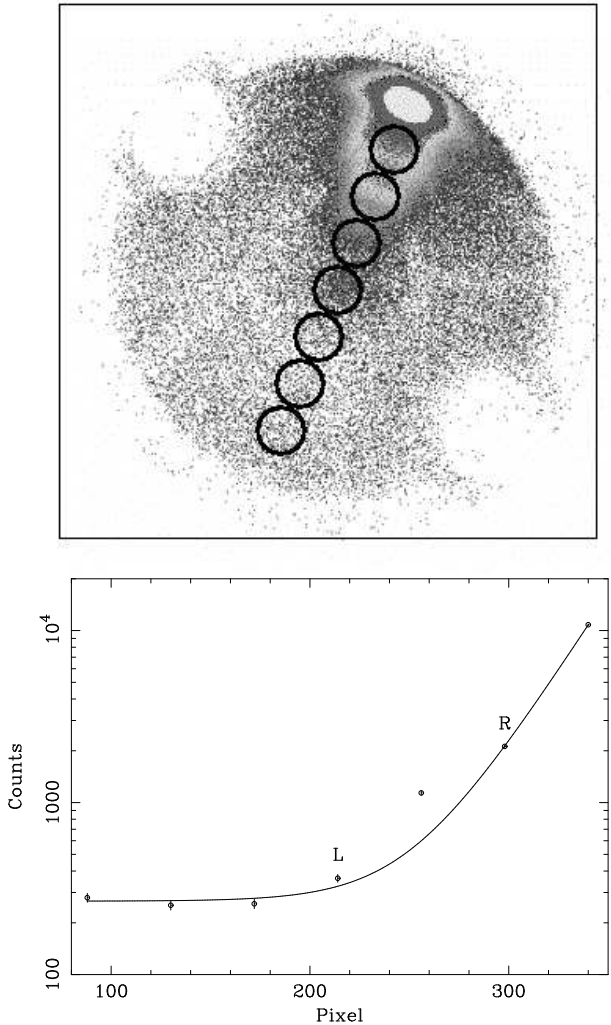,width=8cm,angle=-0,clip=}
\psfig{figure=mineo2490f1b.ps,width=8cm,angle=-90,clip=}
}}
\caption{PSR~B1706$-$44: upper panel shows the MECS image with the regions used 
to accumulate the background, the pulsar position is at the centre of the image.
Lower panel shows the count profile 
vs the centres' coordinates of the regions measured along  
the pulsar-to-binary direction. Points relative 
to the nearest regions (L,R) are also indicated.}
\end{figure}

These values were fitted using a simple analytical formula 
(an exponential plus a constant)  excluding the region containing  
PSR~B1706$-$44; 
the  value computed at the pulsar position was assumed as the proper background 
estimate.
In the lower panel of Fig. 1, we plotted the best fit model from which the excess 
corresponding to the circular region including the pulsar is clearly evident.
We computed   fits correspondent to four energy bands and found that the 
background at the pulsar spot ($B_p$) can be modeled by a linear combination
of the counts in the two nearest regions ($B_L, B_R$): 
$$
B_p\,=\,B_L + b\,B_R  \,\,
\eqno(1)
$$
with the coefficients $b$ practically independent of
 energy. We used $b$=0.13 to estimate the background in the energy 
bins considered for 
the spectral analysis.
\\
The same procedure applied to the LECS data, which have   poorer 
statistics and a wider PSF, shows a detectable signal only in the energy 
channels  lower than 1.5 keV.
We therefore considered only  these photons and included them in a single bin 
from 0.1 to 1.5 keV.

\section{Timing analysis}
We  searched for pulsed emission  in the MECS from the sources
using the Z$^2$ statistics with one and 
two harmonics.
No statistically significant signal was detected for PSR~B0656$+$14 
and PSR~B1706$-$44, even considering different energy ranges.
\\
 A 2 $\sigma$ signal is present at the radio frequence in the LECS data 
of PSR~B0656$+$14 below 2 keV, as expected from the pulsed fraction 
detected by Finley et al. (1992), while
the 2 $\sigma$ upper limit of the pulsed fraction in the energy band 2--10 keV
is 45\%.

We stress that a low pulsed signal such that measured by Gotthelf, Halpern 
\& Dodson (2002)  in PSR~B1706$-$44 emission cannot be detected in the BeppoSAX images, 
where in addition to the contribution of the (unresolved) nebula, 
there is also the local strong background level due to the near LMXB.  

A positive result was indeed obtained for PSR~B1055$-$52.
The Z$^2$ plot gave a peak at the frequency $\nu=5.073285 \pm 0.000005$ Hz,
compatible with that expected by the extrapolation of the ephemeris given 
by Thompson et al. (1999) equal to 5.07328150 Hz. 
Because of its rather low flux we limited the analysis 
to the LECS and MECS events having an energy smaller than 6.5 keV. 
The statistical significance, in this energy range,
derived by the Z$^2$ test for one harmonic corresponds to a chance probability
of 6 $\times$ 10$^{-6}$  (4.5 gaussian standard deviations)

Pulse profiles in the energy bands (0.1$-$6.5 keV),
(0.1$-$2 keV, LECS) and (2.0$-$6.5 keV, MECS) 
are shown in the three panels of Fig. 2: 
they are roughly sinusoidal, as found in the  ROSAT observations 
(\"Ogelman \& Finley 1993).
The significance  of these pulsed signals was also estimated 
with the Z$^2$ test and resulted 2.4 $\sigma$ (probability 
1.5 $\times$ 10$^{-2}$) for the LECS  data and 2.7 $\sigma$ (probability 
7.3 $\times$ 10$^{-3}$) for the MECS.
 Note also that the phase of the pulsation in the two energy bands 
coincides. This is therefore the first 
detection of a pulsed X-ray emission from PSR~B1055$-$52
at energies above 2.5 keV.
To estimate  the pulsed fraction  in the whole energy band, we considered 
 the approximate sinusoidal shape of the signal, and fitted to these data 
(upper panel in Fig. 2) a simple sinusoid added to a constant 

$$
C(\phi) =   C_0 + C_1~sin~(\phi - \phi_0) \,\,
\eqno(2)
$$
and evaluated the pulsed fraction as

$$
f =  [N  - n_b (C_0 - C_1)]/ (N - B) \,\,
\eqno(3)
$$
where $N$ is the total number of counts, $B$ the background level and $n_b$ 
the number of bins in the phase histogram.  
The $B$ value was estimated using archive 
blank fields  corrected to match the local background
and resulted equal to 294 counts (with a negligible error)
 while N was 531.
The fit was  statistically quite good 
($C_0$=53.0$\pm$1.6, $C_1$=15.0$\pm$2.3) and the
resulting pulsed fraction was $f$ = 0.64$\pm$0.17.

\begin{figure}
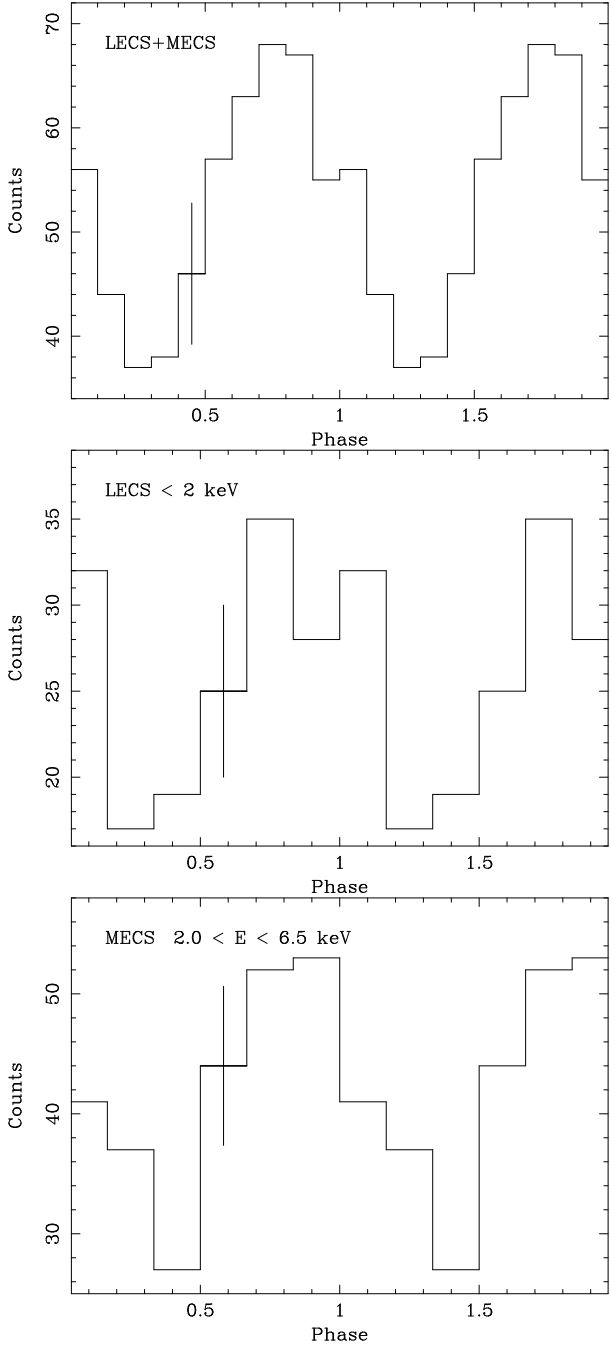

\label{fig1}
\centerline{ 
\vbox{
\psfig{figure=mineo2490f2a.ps,width=8cm,angle=-90,clip=}
\psfig{figure=mineo2490f2b.ps,width=8cm,angle=-90,clip=}
\psfig{figure=mineo2490f2c.ps,width=8cm,angle=-90,clip=}
}}
\caption{Pulse profiles of PSR~B1055$-$52 in different energy bands:
0.1--6.5 keV (upper panel), 0.1--2.0 keV (central panel) and 2.0--6.5 
keV (lower panel)}
\end{figure}

\section{Spectral analysis}
\subsection{PSR~B0656$+$14 }
On the basis of literature results, we used  multi-component spectral 
models to fit the LECS and MECS data. The fits with an absorbed blackbody 
$+$ power law and with two blackbodies did not give acceptable values of the 
reduced $\chi^2$ , as shown in the upper section of Table 2, while a fit with a 
power law plus two blackbodies gave a significant improvement. The 
absorbing column density resulted equal to (3.1 $\pm$ 0.8) $\times$ 10$^{20}$ 
cm$^{-2}$ and the photon index of the power law equal to 2.08 $\pm$ 0.41, while 
the blackbodies' parameters were quite poorly constrained because of the limited 
statistics.    
We therefore added to the BeppoSAX data the  ASCA-GIS 
observation and performed a joint spectral analysis. Again, any attempt to 
obtain a good fit with only two components failed to reach a satisfactory 
reduced  $\chi^2$, as is apparent from the values given in Table 2 (lower section),  
and, as found above, an acceptable fit is reached with two blackbody 
distributions plus a power law. 
The corresponding best fit values of the 
parameters together with 1 $\sigma$ errors are listed in Table 3. The total 
spectrum with the residuals and the unfolded model with the relative components are 
shown in Fig. 3.
The present analysis confirms that the X-ray spectrum of this source is quite 
complex. Our spectral results are generally in  acceptable agreement with 
the previous literature. Typical differences in the blackbody temperatures are of 
the order of  20 \% while a difference of  a factor of two is found between our 
estimate of  $N_H$ is and that of Greiveldinger et al. (1996).
Similar results have been recently obtained by Zavlin, Pavlov \& Halpern 
(2001, as referenced in Becker \& Pavlov 2001) using the same ASCA 
data added to a ROSAT observation.

\begin{table}[h]
\caption{Spectral models fitted to  PSR~B0656$+$14 spectrum.}
\begin{tabular}{lc}
\hline
\multicolumn{2}{l}{BeppoSAX (LECS+MECS)} \\
{\it Model}     &   $\chi^2$ (dof) \\
1) Absorb.  Power Law + Black Body      &  2.98 (13) \\  
2) Absorb.  Black Body + Black Body     &  4.10 (12) \\ 
3) Absorb.  Power Law +  2 Black Body   &  1.29 (11) \\         
\multicolumn{2}{l}{  } \\
\multicolumn{2}{l}{BeppoSAX (LECS+MECS) + ASCA (GIS) }\\
{\it Model}     &   $\chi^2$ (dof) \\
1) Absorb.  Power Law + Black Body      &  1.60 (57) \\  
2) Absorb.  Black Body + Black Body     &  2.40 (56) \\ 
3) Absorb.  Power Law +  2 Black Body   &  0.97 (55) \\         
\hline
\end{tabular}
\end{table}

\begin{table}[h]
\caption{Spectral Parameters for  PSR~B0656$+$14 (model 3).}
\begin{center}
\begin{tabular}{ll}
\hline
\multicolumn{1}{c}{Parameter (unit)} & \multicolumn{1}{c}{Value}  \\
$N_{\rm H}$  (10$^{20}$  cm$^{-2}$)           &     3.4$\pm$1.1\\
kT$_{1}$      (keV)                    &     (5.89$\pm$0.48)$\times$10$^{-2}
$\\
$f_{1}^{a}$  (erg cm$^{-2}$ s$^{-1}$)  &    (2.67$\pm$f 0.16)$\times$10$^{-11}
$\\
kT$_{2}$  (keV)                        &     0.12 $\pm$0.01 \\
$f_{2}^{a}$  (erg cm$^{-2}$ s$^{-1}$)      & (3.49$\pm$ 0.25)$\times$10$^{-12}
$\\
Photon Index                           &     2.10 $\pm$0.23 \\
PL  Norm.                              &     (6.61$\pm$0.86)$\times$10$^{-5}
$\\
\hline
\multicolumn{2}{l} {$^{a}$ unabsorbed flux in the 0.1--2 keV band}
\end{tabular}
\end{center}
\end{table}

\begin{figure}
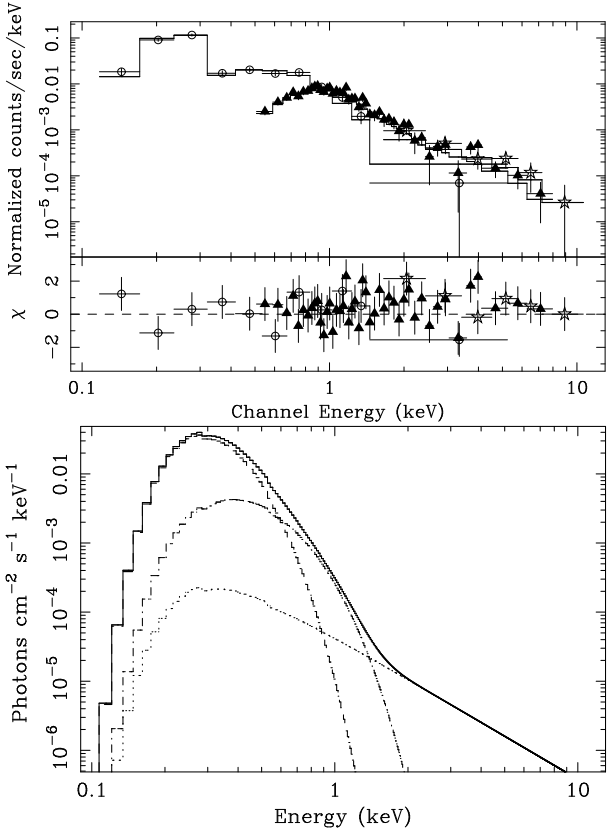

\label{fig2}
\centerline{ 
\vbox{
\psfig{figure=mineo2490f3a.ps,width=8cm,angle=-90,clip=}
\psfig{figure=mineo2490f3b.ps,width=8cm,angle=-90,clip=}
}}
\caption{PSR~B0656$+$14: Spectral fit with model 3 of the LECS 
(open circles), MECS (stars) and ASCA GIS (triangles) data (upper panel).
 Unfolded model with the relative components (lower panel)}
\end{figure}

\subsection{PSR~B1055$-$52}
The rather low intensity of the pulsed signal and the consequent 
difficulty  in the estimation of the unpulsed level did not allow us to perform 
the spectral analysis of the pulsed emission. 
We therefore carried out some fits of spectral models on the total signal. 
A  blackbody law failed to give an acceptable fit ($\chi^2_r$=4.21, 7 d.o.f.), 
being unable to match the flux above a few keV, also an absorbed power law 
gave a $\chi^2_r$ of 1.86 (7 d.o.f.) corresponding to a chance probability 
of about 8\%. Although this fit cannot be completely rejected, we used  
various two--component models: two blackbodies and a blackbody plus a 
power law. 
Both these models  better result than the two others and gave 
$\chi^2_r$ smaller than unity. The estimates of parameters, however, were
rather uncertain, in particular for the absorbing column we obtained
only upper limits. To obtain more information, we performed other
fits with the $N_H$ value fixed at 2 $\times$ 10$^{20}$ cm$^{-2}$,
compatible with the estimate given by Greiveldinger et al. (1996). 
Best fit values with 1 $\sigma$ errors are listed in Table 4 and the 
spectral distribution
for the blackbody plus a power law is plotted in Fig. 4.

\begin{table}[h]
\caption{Best Fit parameters of the two-component models fitted to
 PSR~B1055$-$52 spectrum
($N_H$=2$\times 10^{20}$ cm$^{-2}$).}
\begin{center}
\begin{tabular}{ll}
\hline
\multicolumn{2}{c}{\it Two blackbodies} \\
\multicolumn{1}{l}{Parameter (unit)} & \multicolumn{1}{c}{Value}  \\
kT$_{1}$      (keV)                    &     (7.8$\pm$0.6)$\times$10$^{-2}$\\
$f_{1}^{a}$  (erg cm$^{-2}$ s$^{-1}$)  &     (2.3$\pm$0.3)$\times$10$^{-12}$\\
kT$_{2}$  (keV)                        &     0.74 $\pm$0.14 \\
$f_{2}^{a}$  (erg cm$^{-2}$ s$^{-1}$)  &     (6.7$\pm$1.6)$\times$10$^{-14}$\\
\multicolumn{2}{c}{  } \\
\multicolumn{2}{c}{  } \\
\multicolumn{2}{c}{\it Blackbody + power law} \\
\multicolumn{1}{l}{Parameter (unit)} & \multicolumn{1}{c}{Value}  \\
kT      (keV)                          &     (7.5$\pm$0.6)$\times$10$^{-2}$\\
$f_{1}^{a}$  (erg cm$^{-2}$ s$^{-1}$)  &     (2.2$\pm$0.3)$\times$10$^{-12}$\\
Photon Index                           &     2.09 $\pm$0.49 \\
PL  Norm.                              &     (3.6$\pm$0.8)$\times$10$^{-5}$
\\
\hline
\multicolumn{2}{l}{$^{a}$ unabsorbed flux in the 0.1--2 keV band}
\end{tabular}
\end{center}
\end{table}

\begin{table}[h]
\caption{X- and $\gamma$-ray fluxes for the three pulsars}
\begin{tabular}{lccc}
\hline
\multicolumn{1}{c}{Pulsar} & \multicolumn{1}{c}{$F_{0.1-2}^{a}$} & 
\multicolumn{1}{c}{$F_{2-10}^{b}$}
&\multicolumn{1}{c}{$F_{\gamma}^{c}$}\\
PSR~B0656$+$14 & 30.5$\pm$0.9  & 1.46$\pm$0.02  & 0.04$\pm$0.01   \\
PSR~B1055$-$52 &  2.4$\pm$0.3  & 0.82$\pm$0.19  & 0.33$\pm$0.03    \\
PSR~B1706$-$44 &  1.3$\pm$0.4  & 1.28$\pm$0.07  & 1.12$\pm$0.06    \\
\hline
\\
\multicolumn{4}{l}{$^{a}$ 0.1--2  keV unabsorbed flux in unit of 10$^{-12}$ erg cm$^{-2}$ s$^{-1}$} \\
\multicolumn{4}{l}{$^{b}$ 2--10  keV unabsorbed flux in unit of 10$^{-13}$ erg cm$^{-2}$ s$^{-1}$} \\
\multicolumn{4}{l}{$^{c}$ E$>$ 100 MeV flux in unit of 10$^{-6}$ erg  cm$^{-2}$ s$^{-1}$}\\
\end{tabular}
\end{table}

\begin{figure}
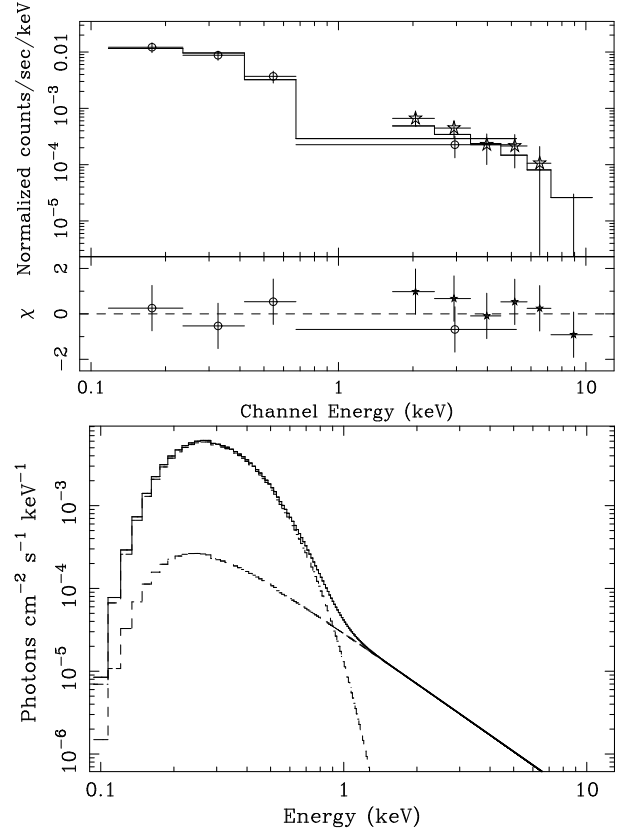

\label{fig2}
\centerline{ 
\vbox{
\psfig{figure=mineo2490f4a.ps,width=8cm,angle=-90,clip=}
\psfig{figure=mineo2490f4b.ps,width=8cm,angle=-90,clip=}
}}
\caption{PSR~B1055$-$52: LECS (open circles) and  MECS (stars) data 
fitted with a blackbody plus 
a power law (upper panel);
 unfolded model with the relative components (lower  panel).}
\end{figure}

\subsection{PSR~B1706$-$44
}

The spectral analysis of  the LECS and MECS data confirmed that the 
spectrum of this source can be well described by a single power law. The fit 
 gave a photon 
index of  1.69 $\pm$ 0.29 and a column density 
$N_{\rm H}$ = (3.7 $\pm$ 1.5) $\times$  10$^{21}$ cm$^{-2}$, with a 
reduced $\chi^2$ = 1.30 for 15 d.o.f. (Fig. 5). 
Fixing the $N_{\rm H}$ value at the 
ROSAT-PSPC result of  5 $\times$ 10$^{21}$ cm$^{-2}$ (Becker et al. 1995), 
we found 
a photon 
index of  1.71 $\pm$ 0.12 (reduced $\chi^2$ = 1.25, 16 d.o.f.), 
in agreement 
with the one reported above. 
To be more confident that our 
result was independent of the local background, we used different estimates 
of its intensity and spectrum. These were derived varying the value of $b$ 
within its uncertainty in the model used to fit the 
LMXB contribution: in all cases we found changes 
of the best fit spectral 
parameter values somewhat less than the statistical uncertainties. 

We are therefore confident that the above result provides the best available 
representation of the actual spectral distribution of PSR~B1706$-$44. 

\begin{figure}
\label{fig3}
\centerline{ 
\vbox{
\psfig{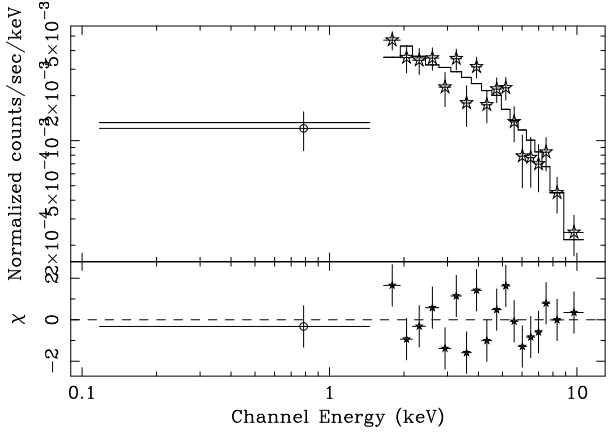}
}}
\caption{PSR~B1706$-$44: LECS (open circles) and  MECS (stars)  
spectra fitted with a power-law  model.}
\end{figure}

\section{Discussion}

The main results of the analysis of the BeppoSAX observations of  
PSR~B0656$+$14 (in this case also joined to a long  ASCA archive observation),
PSR~B1055$-$52 and PSR~B1706$-$44 presented in this contribution can be 
summarized as follows.

\begin{enumerate}
 \item
No pulsation  above 2 keV at the extrapolated radio periods was detected for
PSR~B0656$+$14 and PSR~B1706$-$44, whereas a pulsed signal was clearly
observed for PSR~B1055$-$52  up to the energy band 2.0--6.5 keV with the
relatively high pulsed fraction of 0.64$\pm$0.17, in agreement 
with that given by \"Ogelman \& Finely (1993).
 \item
The spectral distribution of the X-ray emission of PSR~B0656$+$14, derived 
from the combined BeppoSAX and ASCA data, is quite complex. 
Two component models are excluded by the high $\chi^2$ values and an 
acceptable fit is given by at least three components, one power law and two 
blackbodies, as early proposed by Greiveldinger et al. (1996). 
 \item
The X-ray spectrum of PSR~B1055$-$52  is well fitted by a two component
model: a blackbody with a temperature $kT = 0.075$ keV and a power law with 
a photon index of about 2.1. This result does not confirm the very steep 
power law spectrum found by Greiveldinger et al. (1996), while it agrees
with the spectral model given by Wang et al (1998). 

We recall however that a two blackbody model gives also an acceptable fit, but 
a quite high value of $kT = 0.74$ keV is necessary. 
 \item
The X-ray spectrum of PSR~B1706$-$44  is well fitted by a single power 
law with a photon index of 1.65. This value is  
compatible with that connecting the radio to X-ray flux (Finley et al. 
1998) and seems then to confirm that a large fraction of this radiation 
 originates in the compact synchrotron nebula around the pulsar.  
\end{enumerate}

The 0.1--2 keV and 2--10 keV unabsorbed flux values detected by BeppoSAX 
are summarized in Table 5, together with the E$>$100 MeV fluxes (Hartman et al. 
1999) for the three sources. 
Note that the $\gamma$ to X-ray flux ratios are quite different,
suggesting that the radiation mechanisms responsible for these emissions
are not the same in these pulsars. 

 
X-ray emission from young rotation-powered neutron stars is generally
explained either in terms of blackbody radiation from a hot region on 
the star surface
or in terms of non-thermal magnetospheric processes from high-energy
electrons and positrons. Furthermore, a non-thermal component can be 
generated in a surrounding synchrotron nebula. 
Spectral and phase distributions are useful to distinguish between these 
different kinds of emission mechanisms. 

For PSR~B1706$-44$, the intense local background and the presence of a
nebula do not allow us to detect pulsation. According to the recent
results of Gotthelf, Halpern \& Dodson (2002) the dominant component 
originates in the nebula and has a spectral index of 1.34(+0.34,$-$0.20),
statistically compatible with that found in our analysis.
These authors  also found a blackbody component at a temperature of about
1.7 $\times$ 10$^{6}$ K originating from the neutron star, but it is not
necessary to fit our data.

 The complex spectrum of PSR~B0656$+$14 requires at least three components,
two of them having blackbody energy distributions at temperatures different
by a factor of 2. From the measured fluxes of these components, and assuming
a distance for the pulsar, one can estimate some interesting quantities like
the area of the emitting surface and, when this coincides with the entire
neutron star, the stellar radius. Indicating the measured bolometric blackbody 
flux with $F$, the area $A$ of the emitting region is given by

$$
A = (4 \pi d^2 F)/(\sigma_{SB} T^4) \,\,
\eqno(4)
$$
where $d$ is the distance to the source and $\sigma_{SB}$ the Stefan-Boltzmann constant.
For two blackbody components the ratio between their areas is:

$$
A_1/A_2 = (F_1/F_2)\,(T_2/T_1)^4 \,\,
\eqno(5),
$$
while the stellar radius $R_*$ is given by:

$$
R_* = (F/\sigma_{SB})^{1/2}\,(d/T^2) \,\,
\eqno(6).
$$

Bolometric fluxes of the soft (1) and hard (2) components can be derived from 
the (0.1$-2$ keV) fluxes given in Table 3, by applying the proper correction
factors derived from the integrals of the planckian law, which are 1.144 and 1.022
for the softer and harder components, respectively. We found then
$A_2/A_1 = (6.8 \pm 3.7)\times10^{-3}$, coincident with the result of 
Greiveldinger et al. (1996). Such a small value implies that the region of the stellar 
surface emitting the hotter radiation is much less extended than the softer one. 
The latter can be associated with the cooling emission from the entire neutron star 
and Eq.(6) can be used to estimate $R_*$, which for a distance $d$ = 0.5 kpc, 
is 24$\pm$4 km. 
This value is somewhat larger than the one of Greiveldinger et al. (1996), but 
agrees within its satistical uncertainty. Our results therefore substantially
confirm the previous picture of this pulsar.

We stress that the discrepancy, already noted by Greiveldinger et al. (1996), 
between the extent of a pure dipolar polar cap containing the open field lines, 
likely heated by the impact of relativistic particles, and that of the hotter 
region is still present. 
The ratio between the dipolar polar cap area to that of the neutron star surface 
is given by $(\Omega R_*/4c)$ which, considering also a factor of 2 if the spin 
axis is inclined enough to see both polar caps, is equal to $\sim 10^{-4}$, more 
than one order of magnitude smaller than the value given above, 
even taking into account the large uncertainty of the $A_2/A_1$ estimate. 
A possible explanation has been proposed by Wang et al. (1998), who showed
that a hot cap larger than the dipolar one can be due to a relevant distorsion of
the field lines near the star surface produced by the different motion of inner and
outer vortices in the superconducting star core. They give also an estimate of the
cap temperature of $\sim 2\times10^6$ K in good agreement with our findings.

The nature of the X-ray emission from PSR~B1055$-$52 is also rather complex.
A blackbody component with a temperature of about 9 $\times$ 10$^{5}$ K is well
established
 but with flux lower than that reported by Greiveldinger et al. (1996). 
Assuming that it is emitted from the entire surface, it is possible to estimate a
stellar radius $R_* = (8.3\pm1.4) (d/1 kpc)$ km, about one half the
estimate by Greiveldinger et al. (1996) but  compatible with that expected
for a neutron star, unless the actual source distance would be 
somewhat smaller than 1 kpc.

Our results show also that the pulsed emission is detectable at energies higher 
than 2 keV. This finding does not match with a two blackbody model, although 
statistically acceptable,
because it would imply a second emission region at a temperature as high as 
8 $\times$ 10$^{6}$ K, about a factor of 6 larger than that found for 
PSR~B0656$+$14.
The same sinusoidal X-ray pulse shape and phase is observed at energies 
lower and higher than 2 keV, suggesting a common emission site. We recall
that this shape is quite different from that observed
in the EGRET range, which has two peaks with a phase separation of 0.2
(Thompson et al. 1999).
At variance, a non-thermal emission would be more naturally explained by the
model prosed by Wang et al. (1998), where X rays are synchrotron photons 
emitted by secondary electrons produced in the surrounding of the neutron star by
pair absorption of primary curvature $\gamma$-rays from inward-moving 
high energy particles from the outer gap. In this case a relatively broad
pulse profile and pulsed fraction greater than $\sim$0.5 are expected; furthermore,
the spectrum should have a photon index of 1.5. Our results agree 
with these requirements, although the observed index is steeper than the model 
value, but the uncertainty is large enough to make them compatible.

\begin{acknowledgements}
The authors are grateful to Bruno Sacco for his helpful comments. 
 They  are also grateful to the referee, J. Halpern, for the useful
comments. This work
has been partially supported by the Italian Space Agency (ASI). 
\end{acknowledgements}


\begin{thebibliography}{}

\bibitem[Becker et al., 1995]{bec95}
Becker, W., Brazier, K.T.S., Tr\"umper,J. 1995, A\&A, 298, 528

\bibitem[Becker et al., 1999]{bec99}
Becker, W., Kawai, N., Brinkmann, W., Mignani, R. 1999, A\&A, 352, 532

\bibitem[Becker \& Pavlov]{bec01}
Becker, W., Pavlov, G. 2001, "The Century of Space Science", 
Editors: Johan Bleeker, Johannes Geiss and Martin Huber, 
Published by Kluwer Academic Publishers 

\bibitem[Brinkmann, Oegelman 1987]{brin87}
 Brinkmann, W., \"Ogelman, H. 1987, A\&A, 182, 71
 
\bibitem[ Cheng Helfand 1983]{chen83}
 Cheng, A.F., Helfand, D.J. 1983, ApJ, 271, 271

\bibitem[Cordova et al., 1989]{cor89}
Cordova, F.A., Middleditch, J., Hjellming, R.M., Mason, K.O.  1989, ApJ 345, 
451

\bibitem[Fierro et al., 1993]{fier93}
Fierro, J.M.,  Bertsch, D.L., Brazier, K.T.S., et al. 1993, ApJ, 413, L27

\bibitem[Finley et al., 1992]{fin92}
Finley, J.P., \"Ogelman H., Kizilo\v{g}lu, \"U. 1992, ApJ, 394, L21

\bibitem[Finley et al., 1998]{fin98}
Finley, J.P., Srinivasan, R., Saito, Y., et al. 1998, ApJ, 493, 884

\bibitem[Frail et al., 1994]{fra94}
Frail, D.A., Goss, W.M., Whiteoak, J.B.Z. 1994, ApJ, 437, 781

\bibitem[Giacani et al., 2001]{gia01}
Giacani, E.B., Frail, D.A., Goss, W.M.,
Vieytes, M.  2001, AJ, 121, 3133

\bibitem[Greiveldinger et al., 1996]{grei96}
Greiveldinger, C., Camerini, U., Fry, W., et al. 1996, ApJ, 465, L35

\bibitem[Gotthelf et al., 2002]{gott02}
Gotthelf E.V., Halpern J.P., Dodson R. 2002, ApJ, 567,  L125

\bibitem[Hartman et al. 1999]{har99}
Hartman, R.C., Bertsch, D.L., Bloom, S.D., et al. 1999, ApJS, 123, 79

\bibitem[Johnston et al., 1992]{jon92}
Johnston S., Lyne, A.G., Manchester, R.N., et al. 1992, MNRAS, 225, 401

\bibitem[Kifune et al. 1995]{kif95}
Kifune T., Tanimori, T., Ogio, S., et al. 1995, ApJ, 438, 91

\bibitem[Manchester et al., 1978]{man78} 
Manchester R.N., Lyne, A.G., Taylor, J.H., et al. 1978, MNRAS, 185, 
409

\bibitem[Ogelmanfinley, 1993]{oge93}
\"Ogelman, H., Finley, J.P. 1993, ApJ, 413, L31

\bibitem[Pavlov et al., 1997]{pav97}
Pavlov, G.G., Welty, A.D., Cordova, F.A. 1997, ApJ, 489, L75

\bibitem[Ray et 
al. 1999]{ray99}
Ray,A., Harding A.K:, Strickman M. 1999, ApJ 513, 919

\bibitem[Ramanamurthy et al., 1996]{rama96}
Ramanamurthy, P.V., Fichtel, C.E., Kniffen, D.A., et al. 
1996, ApJ, 458, 755

\bibitem[Shearer et al., 1997]{shea97}
Shearer, A., Redfern, R.M., Gorman, G., et al. 1997, ApJ, 
487, L181 

\bibitem[Swanenburg et al. 1981]{swan81}
Swanenburg B.N., Bennett, K., Bignami, G.F. et al. 1981, ApJ, 243, 
69

\bibitem[Thompson et al., 1992]{thom92}
Thompson D.J., Arzoumanian, Z.,  Bertsch, D.L., et al., 1992, Nature 359, 615

\bibitem[Thompson et al., 1999]{thom99}
Thompson D.J., Bailes M., Bertsch D.L. et al. 1999, ApJ, 
513, 297

\bibitem[Vaughan and Large  1972]{vaug729}
Vaughan, A.E., Large,M.I.  1972, MNRAS,
156, 27

\bibitem[Wang et al, 1998]{wang98}
Wang, F.Y.-H., Ruderman, M., Halpern, J., et al. 1998, ApJ, 
498, 373

\bibitem[Zavlin et al.]{zavlin01}
Zavlin, V., Pavlov, G.G., Halpern, J. 2001, preprint

\end{thebibliography}
\end{document}